\documentclass[12pt]{article}
\usepackage{epsf}

\begin{document}

\title{Empiric Models of the Earth's Free Core Nutation}
\author{Zinovy Malkin \\
  Central Astronomical Observatory at Pulkovo RAS, \\ Pulkovskoe~Ch. 65, St.~Petersburg, 196140 Russia \\ malkin@gao.spb.ru}
\date{~}
\maketitle

\begin{abstract}
Free core nutation (FCN) is the main factor that limits the accuracy of the modeling of the motion
of Earth's rotational axis in the celestial coordinate system. Several FCN models have been proposed. A comparative
analysis is made of the known models including the model proposed by the author. The use of the FCN
model is shown to substantially increase the accuracy of the modeling of Earth's rotation. Furthermore, the FCN
component extracted from the observed motion of Earth's rotational axis is an important source for the study
of the shape and rotation of the Earth's core. A comparison of different FCN models has shown that the proposed
model is better than other models if used to extract the geophysical signal (the amplitude and phase of
FCN) from observational data.
\end{abstract}

\section{Introduction}

The precession-nutation motion of the Earth's rotational
axis in space, or, more precisely, in a celestial or
inertial (quasi-inertial) coordinate system, is one of the
main components of Earth's rotation. From a practical
viewpoint, it is used in the procedure of the transformation
of coordinates and vectors of terrestrial and celestial
objects between the coordinate system tied to
Earth's body and an inertial coordinate system, as
required when one solves any astronomical, geophysical,
or navigation task involving observations of celestial
bodies from Earth's surface and observations of terrestrial
objects from space. The motion of the axis of
Earth's rotation in space is described by the theory of
precession-nutation, which is now accurate to 0.1 milliarcseconds (mas)
(IAU2000A model recommended by the International
Astronomical Union). However, a comparison of this
theory to very long baseline interferometry (VLBI)
observations reveals periodic components with amplitudes
as high as 0.3 mas. The main component is due to
the free nutation of Earth's liquid core (FCN, free core
nutation) with a nominal period of about 430 sidereal
days. Figure 1 shows the differences between the
observed coordinates X and Y of the celestial pole and
the position of the pole determined from the IAU2000A
theory of nutation according to data from the International
VLBI Service for Geodesy and Astrometry (IVS)
(Schl\"uter et al., 2002). Unlike the lunisolar and planetary
terms of the IAU2000A theory, this component of
Earth's rotation behaves irregularly and cannot be predicted
theoretically for any sufficiently long time interval
at the required level of accuracy.

\begin{figure}
\centering
\hbox{
\epsfclipon \epsfxsize=0.5\hsize \epsffile{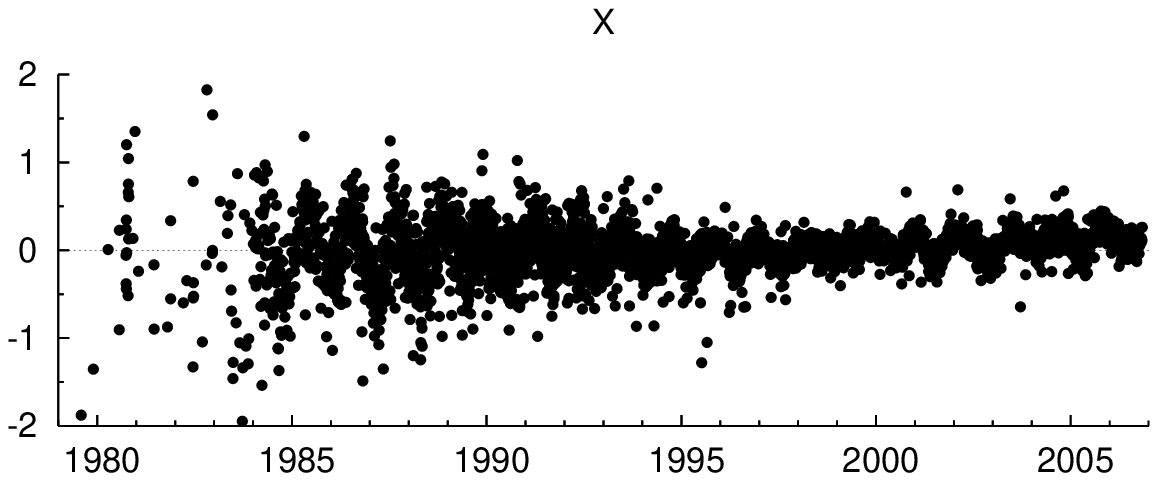}
\epsfclipon \epsfxsize=0.5\hsize \epsffile{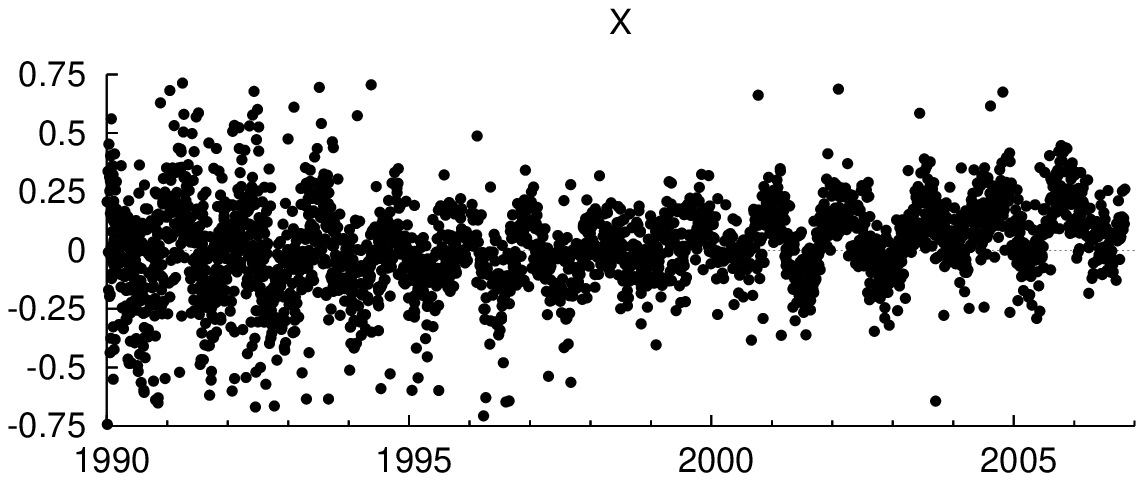}
}
\hbox{
\epsfclipon \epsfxsize=0.5\hsize \epsffile{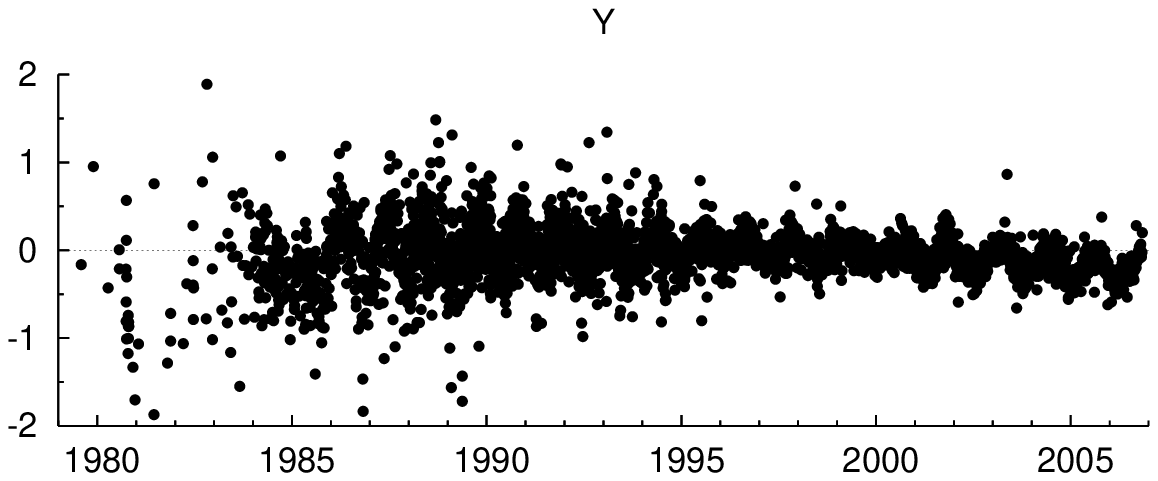}
\epsfclipon \epsfxsize=0.5\hsize \epsffile{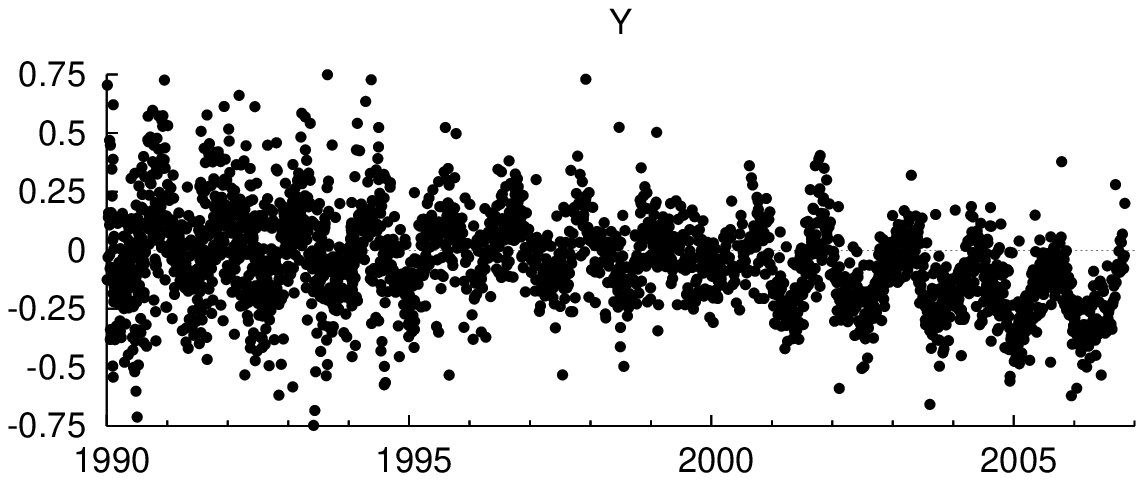}
}
\caption{Differences between the observed coordinates of the celestial pole and its position determined from the IAU2000A theory:
the entire observation period (right) and the period from 1990 onward (left). Unit: mas.}
\label{fig:observ}
\end{figure}

Free nutation of Earth's liquid core was predicted as
early as more than a century ago as one of Earth's rotational
eigenmodes. The frequency of free core nutation
is a fundamental quantity, which appears in the equations of the transfer function that relates the amplitude
of nutation oscillations of the absolutely solid Earth and
the real Earth. The FCN period and amplitude depend
on a number of parameters of Earth's internal structure,
such as the constitution and dynamic flattening of the
core, its moments of inertia, the differential rotation of
the core and mantle, and the dynamic interaction of
Earth shells and their rheological characteristics
(Dehant and Mathews, 2003; Brzezi\'nski, 1996; 2005;
Krasinsky, 2006). However, our knowledge of all these
factors is so far insufficient to construct an FCN model
of sufficient accuracy. Therefore, observational data
and empirical models based on these data provide
important information for refining the parameters of
theoretical models, putting the further refinement of
FCN models at the top of the agenda. For example, the
IAU1980 nutation theory developed before the beginning
of regular VLBI observations used a hydrostatically
equilibrium model of the liquid core, which predicted
an FCN period of 460 d, which differs significantly
from the modern value of 430 d. On the other
hand, the improved accuracy of the empirical models
increases the accuracy of the computation of the nutation
motion of Earth's rotational axis, which is important
for many practical applications. Finally, the need
for highly accurate forecasts of the nutation angles,
which are necessary for many practical applications, is
a factor of no minor importance that stimulates efforts
in the development of FCN models.

All FCN models that we consider in this paper are
based on the analysis of the differences between the
coordinates of the celestial pole determined from VLBI
observations and the coordinates determined from the
IAU2000A theory. This approach assumes that these
differences are mostly due to the contribution of FCN,
which is sufficiently justified by the fact that this
component dominates in the part of Earth's rotational
motion that is not modeled by the IAU2000A theory. At
the same time, the resonance effect of other nutation
terms with periods close to that of FCN may affect the
FCN parameters determined from observations (Brzezi\'nski,
1996; Vondrak et al., 2005). This effect must be
taken into account in the geophysical interpretation of
observational data, but can also be incorporated into the
empirical model intended for practical applications.

In this paper, we consider the three most well-known empirical FCN models
(Herring et al., 2002; Malkin, 2004; Lambert, 2009).
We are the first to compare models from the viewpoint
of how they represent the observed variations of
the FCN amplitude and phase. The results of our comparison
have shown that our model is superior to other
models in this respect. Our comparative analysis also
covers a number of FCN models obtained by simple
smoothing of observational data. We found this procedure
to ensure minimum residuals in the observations
of nutation angles.

\section{Empirical FCN models}

Herring proposed the first FCN model widely used
in practice. Successive versions of this model, which
differed in that they included new observational data,
were incorporated into the KSV and MHB2000 nutation
models (Herring et al., 2002).
The FCN contribution to the coordinates of the celestial pole in the model
considered can be computed as follows:
\begin{equation}\begin{array}{l}
dX = -A_1 \sin \varphi + A_2 \cos \varphi \,, \\
dY = -A_1 \cos \varphi - A_2 \sin \varphi \,,
\end{array}\end{equation}
where $\varphi = -2\pi r(1 + f_0)t$, and $f_0 = -1.00231810920$ is the
FCN frequency, which corresponds to the period of
431.39 sidereal days; $r = 1.002737909$ is the coefficient of
transformation from the mean to sidereal time; $t$ is the
epoch in Julian days counted from the standard epoch
of $t_0$ = J2000.0 = 2451545. The initial amplitudes $A_1$
and $A_2$ were computed from an analysis of the differences
between the observed and theoretical nutation
angles at the epochs of 1979.0, 1984.0-2000.0 with a
two-year step, and 2001.41 (the last epoch after which
the model is not supported). The analysis was performed
on six successive time intervals independently
for both components of nutation (Herring et al., 2002).
In the case of the practical application of this model, the
amplitudes are linearly interpolated to the necessary
epoch. We refer to this model as MHB.

Malkin proposed another FCN model in 2003 (Malkin, 2004).
We denote it as ZM1. According to this model,
the FCN contribution can be computed using the following formulas:
\begin{equation}
dX = A(t) \sin \Phi(t) \,, \quad dY = A(t) \cos \Phi(t) \,,
\end{equation}
where $A(t)$ and $\Phi(t)$ are the amplitude and phase of
FCN, respectively. The variable FCN amplitude can be
computed from the differences between the observed
and theoretical nutation angles:
\begin{equation}
A(t) = \sqrt{dX^2+dY^2} \,.
\end{equation}

The initial differences are first smoothed using a
Gaussian bandpass filter with a central frequency close
to f0, which allows cutting off high- and low-frequency
components lying outside the FCN frequency band.
Smoothing is performed simultaneously with interpolation
to equidistant epochs, usually with a ten-day step,
to simplify and speed up further computations. Note
that Malkin and Terentev (2003) compared the results
of computations made with the initial and smoothed
differences and found no appreciable differences
between them.

We compute the variable phase $\Phi(t)$ as follows. We
first perform the wavelet analysis of the differences
between the observed and theoretical nutation angles to
determine the time variation $\omega(t)$ of the FCN frequency.
We then determine the phase of FCN by integrating the
frequency:
\begin{equation}
\Phi(t) = \int\limits_{t_0}^t \omega dt + \Phi_0 \,,
\label{eq:phase2}
\end{equation}
where $\Phi_0$ is the initial phase at the J2000.0 epoch. The
phase is computed for the same epochs for which the
amplitude has been determined. We finally compute the
FCN contribution using formula (2). Note that a characteristic
feature of the wavelet analysis is the edge
effect, which distorts the data at the extreme subintervals
of the time interval considered. Malkin and Terentev
(2003) and Shirai et al. (2004) analyzed this
effect for the case of FCN. Therefore, although we use
all observations made between 1979-2006 to construct
our model, we give the final series only for the period
1984.0-2005.0.

In 2004, Lambert proposed an FCN model based on principles similar to those of the MHB model
(McCarthy, 2005; Lambert, 2009).
We denote this model as SL. The SL model,
unlike the MHB model, analyzes the differences
between the observed and theoretical nutation angles
strictly by two-year intervals, uses both components of
nutation simultaneously, and, hence, finds the FCN
contribution using the least-squares method in the complex
form:
\begin{equation}
dX + idY = A \exp (i \omega_0 t) + X_0 + iY_0 \,,
\end{equation}
where $\omega_0$ is the circular FCN frequency, which corresponds
to a period of --430.23 solar days (which is equal
to the period of the MHB model), and $X_0$ and $Y_0$ are the
displacements. As a result, formulas to allow for FCN
are derived that are similar to those of the MHB model.
Unlike the MHB model, the SL model is continuously
updated, including its forecast (Lambert, 2009).
The data series begins at 1984.0.

The wavelet analysis of the differences between the
observed and model nutation angles plays an important
part in the ZM1 model. Two comments can be made in
this connection. From the viewpoint of the construction
of the model, the choice of the type and parameters of
the wavelet and the degree of smoothing of the initial
differences are factors that can be varied to achieve the
result that best fits the available observations. Here, we
consider a variant of the ZM1 model, which yields
results that are close to those for the MHB and SL models
as shown below. The second comment concerns the
geophysical interpretation of the model. From the
mathematical viewpoint, the phase variation obtained
from the analysis can be a result of the variation of the
period of nutation oscillations. Strictly speaking, we
have the variations of both the period and phase that can
be separated only by invoking a geophysical analysis. A
number of authors, e.g., Hinderer et al. (2000) and
Zharov (2005), show that the FCN period remains constant
within $\pm$2 d. Therefore, the dependence found is
most likely due to phase variations, which can be correlated
with other geophysical observations. Shirai et al.
(2005) give an example of such a correlation. Thus, the
variations of the amplitude and phase determined based
on empirical models provide material of greatest interest
for further geophysical interpretation.

\section{Comparison of the models}

We compared the FCN models described above
based on the following two criteria. In the first test, we
compared the models using the reduction of residuals in
the observations of the nutation angles. In the second
test, we compared the quality of the representation of
the temporal variations of the FCN amplitude and
phase.

For completeness, we also included into our list of
models to be compared the FCN series obtained by simple
smoothing of the observed differences between the
observed and model nutation angles. This is actually
the series (we denote it as ZM2) obtained at the first
stage of the construction of the ZM1 model. Although
it is not a model in the true sense of the word, it is nevertheless
a good supplement to the IAU2000A nutation
model and allows the accuracy of the modeling of nutation
to be substantially improved in the case of the
transformation between terrestrial and celestial coordinate
systems. It is also easy to forecast both forward
and backward and can, therefore, be applied for operational
and forecasting tasks and for processing old
observations. This series is currently computed for the
period from 1976 to 2010 and is permanently updated.

We test the reduction of observational residuals in
the case of the use of FCN models by analyzing their
spectrum and computing their root-mean-square value.
Figure~2 shows the spectrum of residuals before and
after the application of the ZM1 model. As is evident
from the figure, the spectral peak corresponding to the
frequency of this nutation oscillation virtually disappears
after the application of the FCN model. Other
FCN models described in this paper yield similar
results.

\begin{figure}
\centering
\hbox{
\epsfclipon \epsfxsize=0.5\hsize \epsffile{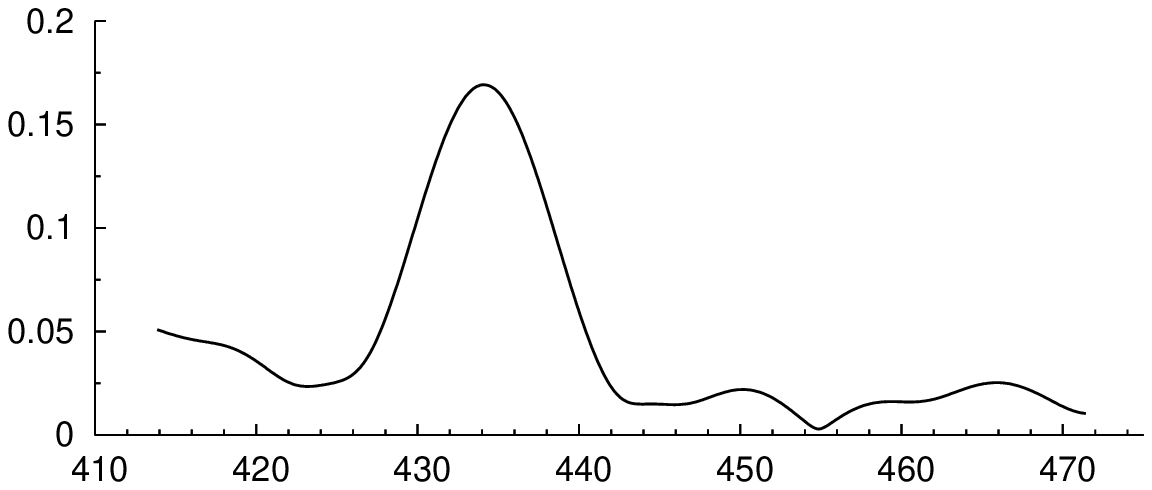}
\epsfclipon \epsfxsize=0.5\hsize \epsffile{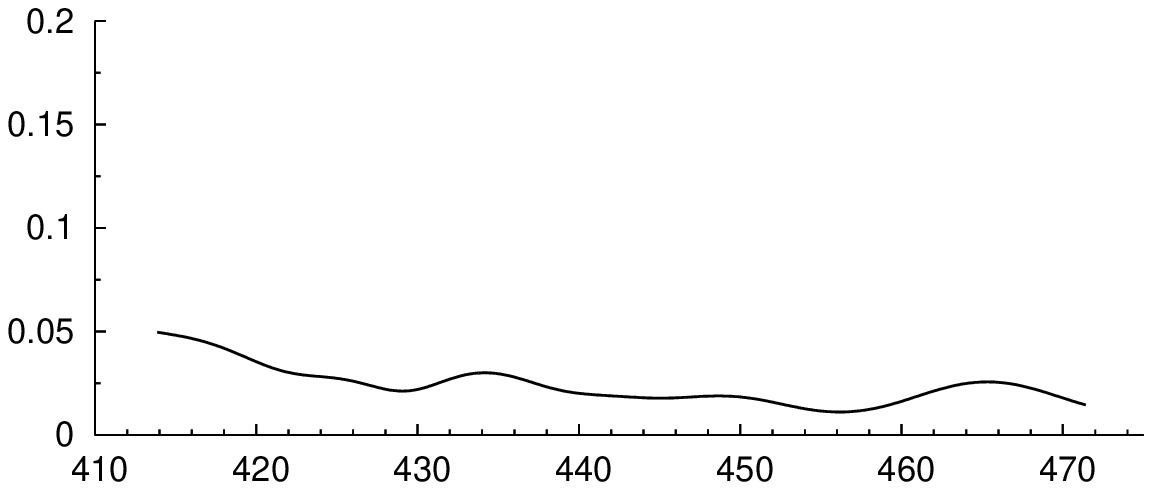}
}
\caption{The spectrum of residuals before (left) and after (right)
correction for the FCN model. Unit: mas.}
\label{fig:spectra}
\end{figure}

Table 1 gives the numerical data on the reduction
of residuals after the application of different FCN models.
This table gives the weighted root-mean-square differences
between the nutation angles determined from
VLBI observations (the combined IVS series) and corrected
by the FCN models and the angles given by the
IAU2000A model. The first line of the table gives the
data for all observation sessions from January 1984
through June 2001, i.e., for the maximum interval of
dates covered by all four compared models. The second
row gives the data for 2003-2004, i.e., for the last two
years when models ZM1, SL, and ZM2 were determined.
We computed the root-mean-square differences
in two variants: `as is' and after the elimination of the
constant shift. The difference between these two variants
is almost negligible over the entire observation
interval, but it is sufficiently large for the last two years.
This is explained by the fact that the differences
between the observed and modelled nutation angles
show an appreciable trend in recent years as can be clearly seen in Fig.~1.
Maybe this trend will disappear after regular
use of the new P03 precession model in the VLBI analysis, which was recommended
by the last General Assembly of the International Astronomical Union in August 2006.

It is evident from the data listed in the table that the
trend affects the residuals for the MHB, ZM1, and SL
models, where the trend is eliminated at the stage of the
construction of the models. At the same time, the trend
has virtually no effect on the residuals for the ZM2
model, because this FCN series was computed without
removing the trend. These results lead us to an important
conclusion about the domains for the optimal
application of different models. The first three models
are best suited for the geophysical interpretation of the
observed data on the free nutation of the liquid core,
whereas the ZM2 series yields the most accurate result
in practical computations involving transformations of
the coordinate system.

We now compare the MHB, ZM1, and SL models
from the viewpoint of the most interesting geophysical
data: variations of the FCN amplitude and phase. For
any FCN model given in the form of a series of corrections
to the X and Y coordinates of the celestial pole, the
FCN amplitude can be computed using formula (3),
and the phase can be determined as atan($X/Y$).
Figure~3 shows the results. The data presented show
that all models exhibit similar variations in the amplitude
and phase of FCN, but the variations yielded by
the ZM1 model are smoother and, thus, apparently better
reproduce the real variations in Earth's rotation. It is
evident from the principles of the construction of the
FCN models that the ZM1 model yields almost continuous
in-time determination of FCN parameters with
any preset step in contrast to the previous models,
which determine these parameters with a two-year time
step, resulting in glitches in FCN parameters, which are
immediately apparent in Fig.~3 for the MHB and SL models.

\begin{figure}
\centering
\hbox{
\epsfclipon \epsfxsize=0.5\hsize \epsffile{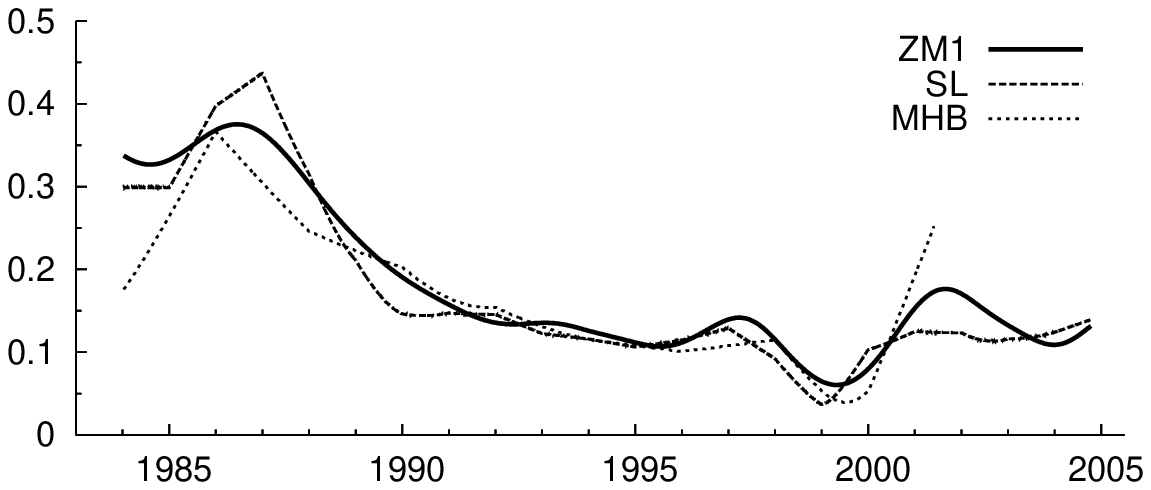}
\epsfclipon \epsfxsize=0.5\hsize \epsffile{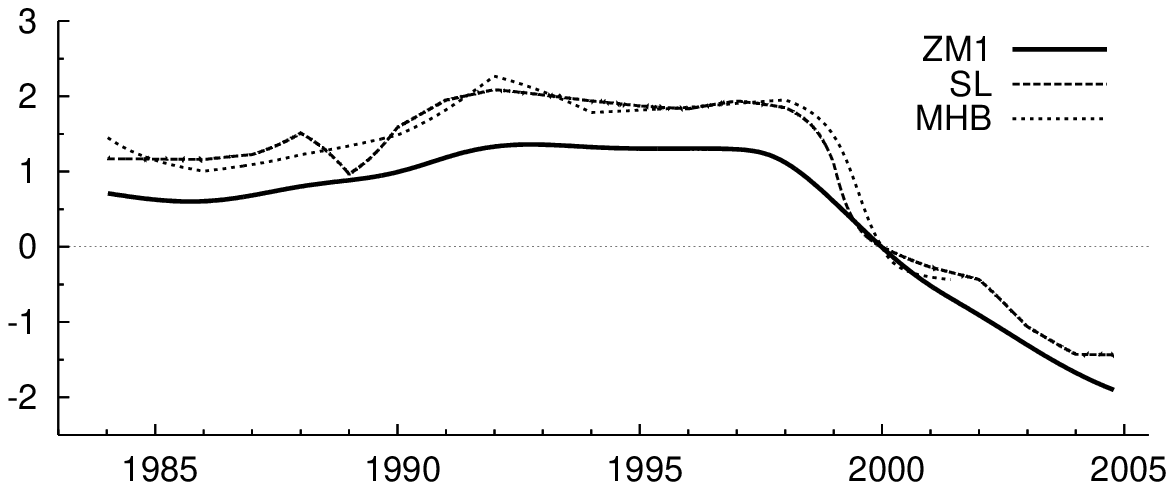}
}
\caption{Variation of the FCN amplitude (left, mas), and phase (right, radians,
linear trend removed) for different FCN models.}
\label{fig:ampl_phase}
\end{figure}

\begin{table}
\centering
\caption{Root-mean-square residuals, mas.}
\label{tab:rms}
\tabcolsep=5pt
\begin{tabular}{|c|ccccc|ccccc|}
\hline
Interval & \multicolumn{5}{|c|}{Raw differences} & \multicolumn{5}{|c|}{After removing trend} \\
\cline{2-11}
of dates & \multicolumn{5}{|c|}{FCN model} & \multicolumn{5}{|c|}{FCN model} \\
\cline{2-11}
& No model & MHB & ZM1 & SL & ZM2 & No model & MHB & ZM1 & SL & ZM2 \\
\hline
1984-2001 & 184 & 155 & 156 & 155 & 143 & 184 & 155 & 156 & 155 & 143 \\
2003-2004 & 156 & --- & 143 & 141 & 89  & 130 & --- & 100 & 102 & 88  \\
\hline
\end{tabular}
\end{table}

\section{Conclusion}

In this paper, we compared several models of the
free core nutation of Earth (FCN). We have shown that
the use of an FCN model substantially improves the
accuracy of the modeling of nutation. A comparison of
different FCN models has shown that our ZM1 model
gives better results if used to extract the geophysical
signal (the amplitude and phase of FCN) from observational
data, whereas the ZM2 series best allows for the
FCN contribution when solving practical tasks involving
the transformation between coordinate systems.

In this paper, we analyzed empirical FCN models
without considering their physical meaning. Many
authors analyzed the physical FCN models and its excitation
based on the available knowledge on Earth's
structure and on the interaction of the outer and inner
Earth layers (Mathews and Shapiro, 1996; Brzezi\'nski
and Petrov, 1999; Shirai and Fukushima, 2001a; 2001b;
Herring et al., 2002; Dehant and Mathews, 2003;
Zharov, 2005; Krasinsky, 2006); however, the analysis
of these models lies beyond the scope of this work.

Note also that, here, we only analyzed the principal
mode of Earth's rotational motion, which corresponds
to the free nutation of the Earth's core with a nominal
period of about 430~d. At the same time, a number of
authors have confidently identified a second oscillation
with a close frequency of about 410--420~d (Malkin and
Terentev, 2003; Schmidt et al., 2005), which can also be
physically explained in terms of a more complex two-
layer model of the liquid core (Krasinsky, 2006; Krasinsky
and Vasilyev, 2006). If the oscillations with a
close frequency are real, they must be incorporated into
all empirical models based on the approximation of the
observed differences of nutation angles by the official
model. At the same time, an analysis of the two-component
FCN model is one of the most interesting directions for further
studies, which will make it possible to
construct a model that would better approximate the
physical properties of Earth and improve the accuracy
of the forecast.

Note, in conclusion, that the ZM1 and ZM2 models
are available at the Internet site of the Pulkovo
Observatory (http://www.gao.spb.ru/english/as/ac\_vlbi/).

\bigskip
\bigskip
\noindent{\Large\bf References}
\bigskip

\leftskip=\parindent
\parindent=-\leftskip

Brzezi\'nski, A., The Free Core Nutation Resonance in Earth
Rotation: Observability and Modelling, In: Proc. Conf. "Present-Day Problems and Methods of
Astrometry and Geodynamics", St. Petersburg, 1996, pp. 328-335.

Brzezi\'nski, A. and Petrov, S., Observational Evidence of the
Free Core Nutation and Its Geophysical Excitation,
Proc. Journ\'ees Syst\`emes de R\'ef\'erence Spatio-Temporels
(1998), Paris, 1999, pp. 169-174.

Brzezi\'nski, A., Chandler Wobble and Free Core Nutation:
Observation, Modeling and Geophysical Interpretation,
Artificial Satellites, J. Planet. Geodesy, 2005, vol. 40,
pp. 21-33.

Dehant, V. and Mathews, P.M., Information About the Core
from Earth Nutation, The Earth's Core: Dynamics,
Structure, Rotation. AGU Geodynamics Series, 2003,
vol. 31. DOI: 10.1029/031GD18

Herring, T.A., Mathews, P.M., and Buffet, B.A., Modelling
of Nutation-Precession: Very Long Baseline Interferometry
Results, J. Geophys. Res., Ser. B, 2002, vol. 107.
No. 4, pp. 2069-2080. DOI:10.1029/2001JB000165

Hinderer, J., Boy, J.P., Gegout, P., et al., Are the Free Core
Nutation Parameters Variable in Time?, Phys. Earth and
Planet. Int, 2000, vol. 117, pp. 37-49.

Krasinsky, G.A., Numerical Theory of Rotation of the
Deformable Earth with the Two-Layer Fluid Core. Pt. 1:
Mathematical Model, Celest. Mech. Dyn. Astron., 2006,
vol. 96, pp. 169-217.

Krasinsky, G.A. and Vasilyev, M.V., Numerical Theory of
Rotation of the Deformable Earth with the Two-Layer
Fluid Core. Part 2: Fitting to VLBI Data, Celest. Mech.
Dyn. Astron., 2006, vol. 96, pp. 219-237.

Lambert S.B. Empirical Model of the Free Core Nutation (Technical Note), 2009. \\
http://callisto.obspm.fr/{\textasciitilde}lambert/fcn/notice.pdf

Malkin, Z.M., Comparison of VLBI Nutation Series with the
IAU2000A Model, Proc. Journ\'ees Syst\`emes de
R\'ef\'erence Spatio-Temporels (2003), St. Petersburg,
2004, pp. 24-31.

Malkin, Z. and Terentev, D., Investigation of the Parameters
of the Free Core Nutation from VLBI Data, Comm. Inst.
Appl. Astron. Russ. Acad. Sci., 2003, no. 149, p. 24;
arXiv:physics/0702152

Mathews, P.M. and Shapiro, I.I., Recent Advances in Nutation
Studies, Proc. Journ\'ees Syst\`emes de R\'ef\'erence Spatio-
Temporels (1995), Warsaw, 1996, pp. 61-66.

McCarthy, D.D., The Free Core Nutation, Proc. Journ\'ees
Syst\`emes de R\'ef\'erence Spatio-Temporels (2004), Paris,
2005, pp. 101-105.

Schl\"uter, W., Himwich, E., Nothnagel, A., et al., IVS and Its
Important Role in the Maintenance of the Global Reference
Systems, Adv. Space Res., 2002, vol. 30, pp. 145-150.

Schmidt, M., Tesmer, V., and Schuh, H., Wavelet Analysis of
VLBI Nutation Series with Respect to FCN, EGU General
Assembly, Geophysical Research Abstracts, 2005,
vol. 7, p. 04555.

Shirai, T. and Fukushima, T., Did Huge Earthquake Excite
Free Core Nutation?, J. Geodetic Soc. Jpn., 2001a,
vol. 47, pp. 198-203.

Shirai, T. and Fukushima, T., Detection of Excitations of Free
Core Nutation of the Earth and Their Concurrence with
Huge Earthquakes, Geophys. Res. Lett., 2001b, vol. 28,
pp. 3553-3556.

Shirai, T., Fukushima, T., and Malkin, Z., Detection of Phase
Jumps of Free Core Nutation of the Earth and Their
Concurrence with Geomagnetic Jerks, 2004, arXiv:physics/0408026.

Shirai, T., Fukushima, T., and Malkin, Z., Detection of Phase
Disturbances of Free Core Nutation of the Earth and
Their Concurrence with Geomagnetic Jerks, Earth Planets
Space, 2005, vol. 57, pp. 151-155.

Vondrak, J., Weber, R., and Ron, C., Free Core Nutation:
Direct Observations and Resonance Effects, Astron.
Astrophys., 2005, vol. 444, pp. 297-303.

Zharov, V.E., Model of the Free Core Nutation for Improvement
of the Earth Nutation, Proc. Journees Systemes de
Reference Spatio-Temporels (2004), Paris, 2005,
pp. 106-109.

\end{document}